\pdfoutput=1
\documentclass[conference]{IEEEtran}
\IEEEoverridecommandlockouts
\usepackage{cite}
\usepackage{amsmath,amssymb,amsfonts}
\usepackage{mathtools}
\usepackage{graphicx}
\usepackage{xcolor}
\usepackage[english]{babel}
\usepackage{newtxtext,newtxmath}

\usepackage{graphicx}
\usepackage{tikz}
\usetikzlibrary{karnaugh}
\usepackage[RPvoltages]{circuitikz} 
\usepackage{amsmath,mathtools}
\usepackage{amsfonts}
\usepackage{graphicx}
\usepackage{algorithm}
\usepackage[noend]{algpseudocode}
\def\BibTeX{{\rm B\kern-.05em{\sc i\kern-.025em b}\kern-.08em
    T\kern-.1667em\lower.7ex\hbox{E}\kern-.125emX}}

\begin{document}

\title{zk-Fabric, a Polylithic Syntax Zero Knowledge Joint Proof System 
}

\author{\IEEEauthorblockN { Sheng Sun}
\IEEEauthorblockA{\textit{Huawei Canada}  rob.sun@huawei.com}
\and
\IEEEauthorblockN{ Dr.Wen Tong (IEEE Fellow) }
\IEEEauthorblockA{\textit{Huawei Canada} tongwen@huawei.com}}

\maketitle

\begin{abstract}
In this paper, we  create a single-use and full syntax zero knowledge proof system, a.k.a zk-Fabric. Comparing with zk-SNARKS and another variant zero knowledge proofing system, zkBOO and it's variant zkBOO++. We present multiple new approaches on how
to use partitioned garbled circuits to achieve a joint zero-knowledge proof system, with the benefits of less overhead and full syntax verification. zk-Fabric
based on partitioned garbled circuits has the advantage of being versatile and single use, meaning it
can be applied to arbitrary circuits with more comprehensive statements, and it can achieve the non-interactivity among all participants. One of the protocols proposed within is used
for creating a new kind of partitioned garbled circuits to match the comprehensive Boolean logical expression with multiple variables,  we use the term "polythitic syntax" to refer to the context based multiple variables in a comprehensive statement. We also designed a joint zero knowledge
proof protocol  that uses partitioned garbled circuits. 
\end{abstract}


\section{Introduction}
In cryptography, “Zero-knowledge” proofs allow one party (the prover) to prove to another (the verifier) that a statement is true, without revealing any information beyond the validity of the statement itself. For example, given the hash of a random number, the prover could convince the verifier that he/she actually owns the number, without revealing what it is, he/she can construct a proof system to let the verifier to be convinced. A zero-knowledge proof convinces a verifier of a statement while revealing nothing but its own validity. Since they were introduced
by Goldwasser, Micali, and Rackoff \cite{c1}, zero-knowledge (ZK)
proofs have found applications in domains as diverse as authentication and signature schemes \cite{c2}, secure computation \cite{c3},
and emerging shield transaction in blockchain technologies \cite{c4}. With the state of art zero knowledge technologies, such as zk-SNARK \cite{c5}, zkBOO \cite{c6} etc, it brings new versatile  functionalities to today's infrastructure.

This means that, given
an interactive proof for any NP-complete problem, one can
construct zero-knowledge proofs or arguments for any NP statement. But existing solutions has the limitations of monolithic state, in another word, the NP statement contains one argument at a time, another limitation are the computational overheads, mainly because many of the early techniques \cite{c5,c6} require many iterations of finding the arithmetic roots of a polynomial equations
to achieve negligible soundness error. The overhead is observed \cite{c7} in the prover work and communication. More recent work
\cite{c6} avoids those issues, but generally entails many expensive cryptographic operations.

zk-Fabric is constructed as a variant of zk-SNARK and zkBoo, a novel form of zero-knowledge cryptography. zk-Fabric inherits the strong privacy assurance with full syntax verification capability,  that multiple variables within a statement can be fully verified  without revealing the true value. The zk-Fabric can be benefited in the Blockchain, with the anonymous verifiers jointly provides the Zero Knowledge Proof.

 In zk-SNARK, the prover and verifier had to endure a heavy trusted setup which consumes large sets of cryptographic primitives and running time \cite{c7}. zkBoo and its derivative,zkBOO++ \cite{c8} have constructed zero knowledge proof using a difference approach without the requirements of trusted setup and not reliant on the arithmetic circuit, instead it's employing the garbled circuit \cite{c9,c10} to construct the zero knowledge proof system. However both systems are built inherently to solve monolithic statement which hinders its practicality to large system, such as contract verification, auditing etc. zk-Fabric is built to overcome the shortcomings of zk-SNARKs and zkBOO with efficient way to produce joint zero-knowledge proofs based on garbled circuit regime. zk-Fabric also maintains what is deemed as important features of online zero knowledge proof system, that are  non-interactive and succinct to publish to a block chain which provides the publicly retrievable information, perfectly for the shield auditing service. 

zk-Fabric also creates the protection mechanisms to prevent the false proofs, known as semi-honest model \cite {c11}, specifically if someone had accessed to the secret randomness used to generate these parameters, they would be able to create false proofs that would look valid to the verifier.  To prevent this from  happening, zk-Fabric generates the public parameters through the partitioned garbled circuits with multi-party settings.

In section 2, a pre-settings of cryptographic notions and functions are introduced
In Section 3 we introduce the zk-Fabric schematics, including the framework which integrate all functional modules together. In Section 4, we introduce the first module whose job is to transform a full semantic statements from a prover into a  polylithic syntax logical expression with Boolean operations, note the word "polylithic" is created in comparison with "monolithic". In section 5, we introduce the second module, which creates partitioned garbled circuits  on the basis of the generated polylithic syntax expression, and prepare the garbled circuit with public cryptographic primitives. In Section 6, we introduce the 1-2 OT (oblivious Transfer) \cite {c13} based OT-aggregator verification protocol, which obtains the zero-knowledge proofs with the partitioned Garbled circuits created by module II in section 6.  
 Lastly, we draw our conclusion and present ideas for future work in section 8.

\section{Preliminaries and Notions}
 In this section, we will provide some insights into the cryptographic concepts employed in our zk-Fabric scheme, in which most of the designs play vital role in building the zero-knowledge proof system. We will also introduce the notions within the paper.

\subsection{Notions}
In our paper, we denote Alice as the garbled circuits sender, and Bob as the garbled circuits receivers. In the joint 
 verification schemes, Charlie, David are also referred as garbled circuits receivers; We also denote $x_i$ as the inputs from sender and receivers; $x_i^j$ is used to denote the $ith$ input $x_i\in(0,1)^\ell$ from $j^{th}$ receiver, in most circumstances, $j=1$ represents the sender, where $\ell$ denotes the length of the bit string; We also denote $C_i$ as the $i^{th}$ garbled circuit constructed in section 3. Similarly, $e_i^j$ denotes the encoding function in a garbled circuit scheme fro $j^{th}$ receiver; $d_j$ denotes the decryption key for $j^{th}$ receiver; $Y_i$ denotes the evaluation output for $i^{th}$ garbled circuit; $Y$ denotes the value of unified Boolean operations on $Y_i$; $y$ denotes the expected value.

\subsection{Yao's Garbled Circuit} 

\noindent A garbled circuit is a method to "encrypt a computation" that reveals only the output of the computation, but reveals nothing about the inputs or any intermediate values. The "circuit" is referred to a combination of logical operations on inputs, and the syntax is expressed as a Boolean circuit, with the Boolean gates, such as (AND, OR, NOT) gates in the circuit. An example of a logical circuit can be illustrated as:

\noindent A classical Yao's "garbling scheme" \cite{c9} consists of:

\begin{itemize}
 \item[1]  (Garbler): comprises of a method to convert a (plain) circuit $C$ into a garbled circuit $\hat{C}$.

 \item[2]  (Encoder) comprises of a method to convert any (plain) input $x$ for the circuit into a garbled input $\hat{x}$. You need the secret randomness that was used to garble the circuit to encode $x$ into $\hat{x}$.

  \item[3]  (Verifier) comprises of a method to take a garbled circuit $\hat{C}$ and garbled input $\hat{x}$ and compute the circuit output $C(x)$. Anyone can do this, you don't have to know $x$ or the secret randomness inside $\hat{C}$ to evaluate and learn $C(x)$.

\end{itemize}  

\noindent The main idea of security is that $\hat{C}$ and $\hat{x}$ together leak no more information than $C(x)$. In particular, they reveal nothing about $x$, yet they allow the computation $C(x)$ to be completed. In many terms, it's often referred as  "Encrypted Computation".

\subsection{Karnaugh Map} 
Karnaugh Maps \cite{c12} offer a graphical method of reducing a digital circuit to its minimum number of gates. The map is a simple table containing 1s and 0s that can express a truth table or complex Boolean expression describing the operation of a digital circuit. The map is then used to work out the minimum number of gates needed, by graphical means rather than by algebra. Karnaugh maps can be used on small circuits having two or three inputs as an alternative to Boolean algebra, and on more complex circuits having up to 6 inputs, it can provide quicker and simpler minimisation than Boolean algebra. 

We employ Karnaugh Maps as the tool to reduce the complex syntax (polylithic) into minimum gate setups, which in turn helps reduces the computational and communication overhead associated with zk-Fabric. 

\subsection{1-2 Oblivious Transfer}
In cryptography, an oblivious transfer (OT) protocol is a type of protocol in which a sender transfers one of potentially many pieces of information to a receiver, but remains oblivious as to what piece (if any) has been transferred.  There are variants of oblivious transfer protocol \cite{c13,c14,c15}, in our paper, we utilize the 1-2 oblivious transfer in our zk-Fabric system. 

In a 1–2 oblivious transfer protocol \cite{c13},   Alice prepares two messages $m_0, m_1 \in (0,1)^\ell$ 
of length $\ell$ and sent them to the receiver, and the receiver has to choose which one of them
is disclosed. The sender does not know, which message $m_b$ is delivered
and the receiver has no possibility to get any information about the other
message $m_1-b$ The protocol consists of at least two messages: Choose
contains the receiver’s choice $b$ and Transfer contains both messages $m_0$
and $m_1$ of which only $m_b$ can be read by the receiver.

\section{zk-Fabric System} 

We will now provide an overview of zk-Fabric and our constructions of this new cryptographic system. zk-Fabric consists of three major components: I) Polylithic Syntax Decomposition; II) Construction of partitioned garbled circuits; III) A Non-interactive OT based Multi-Parties joint Verification scheme.

\subsection{zk-Fabric in a nutshell}
We start with the construction of a overall zk-Fabric system that builds on the partitioned OT scheme. This construction was inspired by the security notions of OT-Combiners 
\cite{c18}. Figure 3 shows an example of 2 polylithic
inputs to be ”blindly” verified by 3 offline verifiers with the
construction of partitioned garbled circuits.

\begin{figure}[ht]
\centering
\includegraphics[width=0.9\columnwidth]{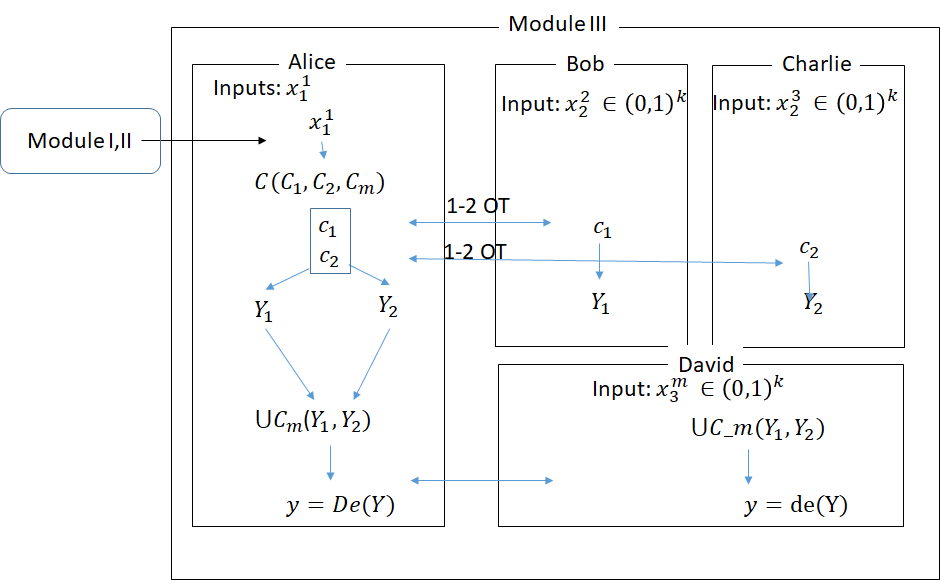}\hspace{2pc}%
\caption{zk-Fabric system }
\end{figure}

Alice bootstraps the zk-Fabric by invoking Module I in section 4 and Module II in section 5 which are to prepare a partitioned garbled circuits for multiple verifiers. The prover Alice distributes the garbled circuits through the Module III in section 6, the extended OT schemes with the verifiers through online systems, such as Blockchain, or web portals that are publicly accessible. Through the public online systems,  the verifiers, Bob and Charlie in this example,  compute the intermediate circuit output $Y_i$ corresponding each partitioned garbled cicuit $C_i$. At the final step, a garbled circuit OT-aggregator David computes the $Y$  and verifies if the $Y=y$. 

In zk-Fabric, we consider the circuit $C$ a garbled circuit evaluation problem. In this problem, the prover Alice constructs a circuit $C$, consisting of Boolean gates operations over finite field $\mathbf{GF(2)}$.  We generalize the $C$ can be partitioned with degree $m$ with the input vector $X$. The $C$ therefore can be represented with a vector $C:(C_1,C_2,...,C_m)$. The evaluation goal is to evaluate $C$ on inputs $X$.In  an non-interactive proof for this problem, the prover sends the outputs vector $Y:(Y_1,Y_2,...)$ of $C$ on input $X$, the verifiers must prove that $Y=C(X)$.

\emph{Definition 2.1: A zk-Fabric scheme syntactically consists of three Modules  I:synGen, II: xgcGen, III: OT-aggregator}.

\begin{itemize}
    \item Module I: $C \leftarrow synGen(1^\lambda)$:\emph{It takes input vector $m:(m_1,m_2,...)$, map the $m$ into hashed vector $X:(x_1,x_2,...)^\lambda \gets hash(m)$ as inputs of the security parameter $1^\lambda$ and outputs a Boolean gates based expressions C. } 
    \item Module II: $(C_1,C_2,...,C_m) \leftarrow xgcGen(C)$: \emph{It takes the Boolean gates expression C into an enumerated length of $m$ circuits $(C_1,C_2,...) $ of garbled circuits  with the native garbled circuit generator algorithms.}
    \item Module III: $Y \leftarrow OT-aggregator(Y_1,Y_2,...Y_{m-1})$: \emph{ It takes inputs from outputs of the partitioned garbled circuits $(Y_1,Y_2,...,Y_{m-1})$, and outputs an aggregated output $Y$. The ultimate verification algorithm is to verify if $Y$ is equal to the expected output for the extended garbled circuit algorithm $xgc$ :} 
 \begin{equation}
   \left\{Y=C(X)\vert y=f(m \oplus x_m^m)\right\} =
    \left\{
    \begin{array}{ll}
     True  & if \ Y=De(y) \\
     False & Otherwise\\
    \end{array} 
    \right.
\end{equation}

   \item[1] An universal Hash Algorithm $x_i \leftarrow h(S_i)$: It takes a common reference string ($crs$)  $S_i$ as inputs and outputs a uniformly distributed digest. The $h$ is an universal hash function.
   \item[2] Extended garbled circuit scheme $gc$: The extended garbled circuit is a four-tuple algorithm ${gc}=(Gb,Enc,De,Ev)$, where $Gb$ is a randomized garbling algorithm that transform $f$ into a triplet ($C_i,e_i,d_i$), the ${C_i}$ is the $i^{th}$ partitioned garbled circuits, $e_i$ is the corresponding encoding information for circuit $C_i$, and $d$ is the corresponding decoding information. $Enc()$ is an encoding algorithm that maps input $X_i$ into garbled input via $X_i=Enc(e_i,x_i)$. $De$ is a decoding algorithm that maps the garbled output $Y_i$ into plaintext output $y_i=De(d_i,Y_i)$. $Ev()$ is the algorithm with input $X_i$ and $F_i$ which generates garbled output $Y_i=Ev(F_i,X_i)$.

   \end{itemize}

\emph{Definition 2.2:  We require that zk-Fabric holds the following security properties}:

\begin{itemize}
    \item  Correctness: \emph{It should hold for all common reference string with input size $\ell$, all garbled circuits $C_i \in[C]$,} 
   
  \begin{equation}
  Pr \left\lbrack y=Y \middle\vert
  \begin{array}{l}
    X(x_1,x_2,...)^\lambda \leftarrow h(s_1,s_2,...)^r\\
    C \leftarrow synGen(X)\\
    C(C_1,C_2,...,C_m) \leftarrow xgcGen(C) \\
    y_i=Y_i, \forall i\in C_i\\ 
    Y_i=gc(C_i), \forall i \in C_i \\
    Y \leftarrow OT-aggregator(Y_1,Y_2,...,Y_{m-1})
  \end{array} \right\rbrack \approx 1
\end{equation}

  \item Privacy: \emph{zk-Fabric should hold the perfect sender and receivers' privacy with the underlying OT scheme's privacy model and the OT-aggregator model. Specifically, for any non-interactive multiple parties oblivious transfer protocol $(S, R_i)$ between a sender $S$ and multiple receivers $R_i$, it satisfies that any of the receiver $R_i$ does not learn any information on the input bits mapped to the corresponding circuit $C_i$, with a negligible probability $\epsilon \approx \frac{1}{2^\ell}$}

\end{itemize}

\section{Module I: Polylithic Syntax Construction}
We start the zk-Fabric protocol at the site of prover Alice who has a composite statement to be verified without revealing the real value.The composite statement can be trivially converted into the regular expressions which represent a set of strings with some tools, i.e intrusion detection systems SNORT. In the Polylithic Syntax construction module, we can employ the regular expression matching techniques to detect a pattern written by regular expressions from the input strings. 

An example of such compound statements could be a simple sentence, such as: "\underline{The car only starts} $\lbrack if \rbrack$ \underline{the "start" button is pressed}  $\lbrack and \rbrack$ \underline{ the brake pedal is pressed}" represents the variables $S_i$ to be verified in operator vector $o$, all the  $\lbrack \cdot \rbrack$ represents logical relationships between variables $v$. Compared with native zero-knowledge proof system which can only process single(monolithic) variable at a time, zk-Fabric is aimed to match the regular expression with patterns and construct corresponding circuits with reduced complexity.

In construction of the logical gates, we utilize the Karnaugh Map technique to reduce the logical expression complexity since the efficiency of the zk-Fabric is depending on the circuit complexity with depth of $\ell$. Due to the limitations of Karnaugh Maps, it's ideal to constrain the zk-Fabric with $\leq6$ inputs in the input vector $S$.  The Karnaugh Map is introduced in section 2. 

W.l.o.g, we can assume a few pre-setting functions existed to assist the constructions of the polylithc syntax conversion. 

\begin{itemize}
\item \emph {Definition 4.1:  there exists a generalized extraction function which can match the compound expression in strings with key variables and the logical relationships expressed in Boolean operators [AND, OR, XOR,etc], we can define the functions as $Extractor_v()$  and $Extractor_o()$, which recognize match the variables and logical operators respectively.}
\item \emph {Definition 4.2:   there exists a generalized regular expression function which can match the regular expression in strings with certain patterns and convert the parameters and patterns into a regular expression, we define the function as Regexp() \cite{c16}.}
\item \emph {Definition 4.3:  there exists a generalized function which can covert a regular expression string into logical circuits \cite{c16}, we can define this function as $CircuitGen()$. }
\item \emph {Definition 4.4:   there exists a generalized Karnaugh Map algorithm which can further reduce the logical gates complexity,and we can define this function as $K-map()$}.
\end{itemize}

We implement a generalized algorithm for Polylithic Syntax generation as  algorithm 1. 

\begin{algorithm}
\caption{Polylithic Syntax Generation Algorithm }\label{euclid}
\begin{algorithmic}
   \Require Composite String \textit{string}
   \Ensure Circuit ( \textit{C})
    $strlen \gets |string|$
    \While {\textit{in strlen}} 
          \State $S' \gets Extractor_v(string)$ 
          \State $O \gets Extractor_o(string)$ \;
          \State $S \gets hash(S')$  \; 
          \EndWhile 
     \State $Exp \gets Regexp (S,O)$  \;
     \State $C' \gets Circuit-Gen(Exp)$ \;
     \State $C \gets K-map (C')$ \;
\end{algorithmic}
\end{algorithm}

\section{Module II: Partitioned Garbled Circuits Construction}
With the inputs a circuit $C$ which represents a list of truth tables along the depth of the logical gates operations, we can implement the garbled circuit scheme with non-interactive commitments to ensure the verifiers can achieve the correctness and privacy in verification. We employ the abstraction of garbling schemes \cite{c17}, which was briefly introduced in section 2. The correctness property of a garbling scheme is defined as, $\forall (C,e,d)$ in the support of $Gc(1^k,c)$ and all inputs $x$, we have $De(d, Ev(C,En(e,x)))=c(x)$, where the $k$ denotes the security parameter.

\subsection{Garbled circuit Representation}
A Boolean circuit can be thought of as a Directed Acyclic Graph (DAG),i.e. a graph with no loops, with each node representing a unit of computation
performing a specific operation op (e.g. AND/XOR). Moreover, we fix all gates to have 2 input wires, a left and a right wire which we denote by $l$ and
$r$, that functions, which given a wire returns its bit value 0 or 1. We denote
the depth of a circuit by $d$, and its width by $n$. While there are many ways
to represent them, a convenient way to think of them is as a $M: d \times n$ matrix,
i.e. each layer $i \in  (1,..,d)$ of a circuit has a fixed width $n$, with each entry being
a gate. We find this representation convenient in defining the composition of layers (to obtain a circuit)

\subsection{Partition Garbled Circuits Scheme }
We would require a partition scheme in our multiple parties OT verification protocol, meaning a $xGC$ can properly partition the garbled circuits $C$ into multiple independent garbled circuits, recorded in a vector $C(C_1,C_2,...)$. Given a yao's circuit $C$ with a matrix of inputs $[x_i]$ and the outputs $[o_i]$ within a truth table, our goal is to securely divide the table into multiple representations of the truth table matrix $T$. To obtain the security properties of such scheme, it can be proved with state-separating proofs \cite{c19}  
and to make cryptographic proofs more suitable for multi-parties verification. The following protocol defines the scheme of partitioned garbled circuit construction. 
\\

\textbf{Partitioned Garbled Circuit Generation Scheme:}

\begin{itemize} 

\item[1] Preparation: we denote the multi-parties Yao's garbled circuit with sorted inputs $x_1,x_2,...,x_n$, we pair two sorted inputs together onto one Boolean gates, if there exists odd number of inputs, two additional auxiliary random value $a_0,a_1 \in \{0,1\}$ will be added. In a trust setup, the auxiliaries will have to destroyed at the "toxic wastes" after being used. We can define a new circuit $C'$ with this preparation:

\begin{equation}
 C'(x_1,...,x_n)= \left\{
\begin{array}{cc}
      C((x_1, x_2),...(x_{n-1}, x_n))  & \textit{if n is even}  \\
      C((x_1, x_2),...(x_n, (a_0 \oplus a_1)) & \textit{if n is odd} \\
\end{array} 
\right. 
\end{equation}

\item[2] Garbled Circuit Construction: for each input pairs $(x_i,x_j)$, where $x_i$ denotes the input from the prover, and the $x_j$ denotes the input from the verifier,  and the wires and internal wires $w$ of the circuit, assign a pair of keys $(k_w^0,k_w^1)$. 
\item[3] Garbled Circuit Construction: for each gate of the circuit, generate 4 ciphertexts which encrypts the corresponding key associated with the output wire according to the truth table of the table $T$,  Figure 4 illustrates the wire assignments and the outputs.
\item[4] Garbled Circuit Construction: for each gate connected to an output wire of the circuit, we encode 0/1 according to the truth table as the Yao's garbled circuits scheme.

 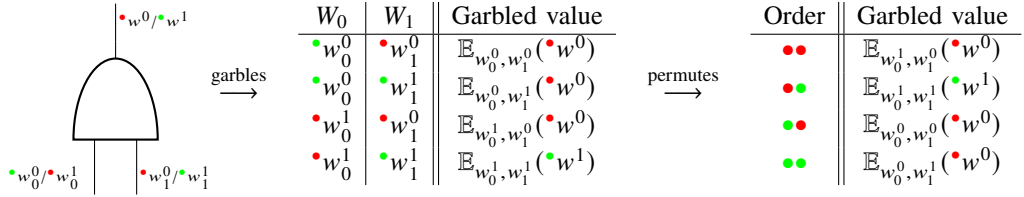
\begin{figure*}[ht]
 \begin{center}

	\begin{tabular}{ccccc}
		\begin{tikzpicture}[baseline=-4ex,inner sep=0pt,font=\tiny,text height=3pt,text width=15pt]
			\draw (0,0) node[rotate=90,and port] (andp) {};
			
			\draw (andp.in 1) --++(0,-0.5) node[pos=0.65,xshift=-25pt]{$
				\prescript{\color{green}\bullet}{}w_0^0 /
				\prescript{\color{red}\bullet}{}w_0^1
				$};
			\draw (andp.in 2) --++(0,-0.5) node[right,pos=0.65,xshift=1pt]{$
				\prescript{\color{red}\bullet}{}w_1^0 /
				\prescript{\color{green}\bullet}{}w_1^1
				$};
			\draw (andp.out) --++(0,0.5) node[right,pos=0.45,xshift=1pt]{$
				\prescript{\color{red}\bullet}{}w^0 /
				\prescript{\color{green}\bullet}{}w^1$};
		\end{tikzpicture}
		&
		$\stackrel{\text{garbles}}{\longrightarrow}$
		&
		$\begin{array}{c|c||c}
		W_0 & W_1 & \text{Garbled value} \\ \hline
		\prescript{\color{green}\bullet}{}w_0^0 & \prescript{\color{red}\bullet}{}w_1^0 &  \mathbb{E}_{w_0^0,w_1^0}(\prescript{\color{red}\bullet}{}w^0) \\
		\prescript{\color{green}\bullet}{}w_0^0 &
		\prescript{\color{green}\bullet}{}w_1^1 & 
		\mathbb{E}_{w_0^0,w_1^1}(\prescript{\color{red}\bullet}{}w^0) \\
		\prescript{\color{red}\bullet}{}w_0^1 &
		\prescript{\color{red}\bullet}{}w_1^0 &
		\mathbb{E}_{w_0^1,w_1^0}(\prescript{\color{red}\bullet}{}w^0)\\
		\prescript{\color{red}\bullet}{}w_0^1 &
		\prescript{\color{green}\bullet}{}w_1^1 &
		\mathbb{E}_{w_0^1,w_1^1}(\prescript{\color{green}\bullet}{}w^1)	\\
		\end{array}$
		& 
		$\stackrel{\text{permutes}}{\longrightarrow}$
		&
		$\begin{array}{c||c}
		\text{Order} & \text{Garbled value} \\ \hline
		{\color{red}\bullet\color{red}\bullet} &
		\mathbb{E}_{w_0^1,w_1^0}(\prescript{\color{red}\bullet}{}w^0)
		\\        
		{\color{red}\bullet\color{green}\bullet} &
		\mathbb{E}_{w_0^1,w_1^1}(\prescript{\color{green}\bullet}{}w^1)	
		\\
		{\color{green}\bullet\color{red}\bullet} &  \mathbb{E}_{w_0^0,w_1^0}(\prescript{\color{red}\bullet}{}w^0) 
		\\
		{\color{green}\bullet\color{green}\bullet}& 
		\mathbb{E}_{w_0^0,w_1^1}(\prescript{\color{red}\bullet}{}w^0) 
		\\
		\end{array}$
	\end{tabular}
	\caption{Yao's Garbled Circuit}
	\end{center}
\end{figure*}

\item[5] Partitioning Garbled Circuit: based on the Truth table $T$, we can partition the circuit matrix $M$ horizontally to the penultimate gate before the last aggregating gates. The partitioning of garbled circuit needs some extra care to maintain the inputs/outputs integrity, we implement the partitioning with $n/1$ (fan-in / fan-out) ratio. Specifically, the $n/1$ scheme requires that the leftmost input gates are partitioned per garbled logical gate, and after the first tier of inputs, the intermediate and last tier gates are aggregated into one garbled circuit. An example of the partitioning the  gates is illustrated as in Figure 5. 
\begin{figure}[ht]
\centering
\includegraphics[width=0.9\columnwidth]{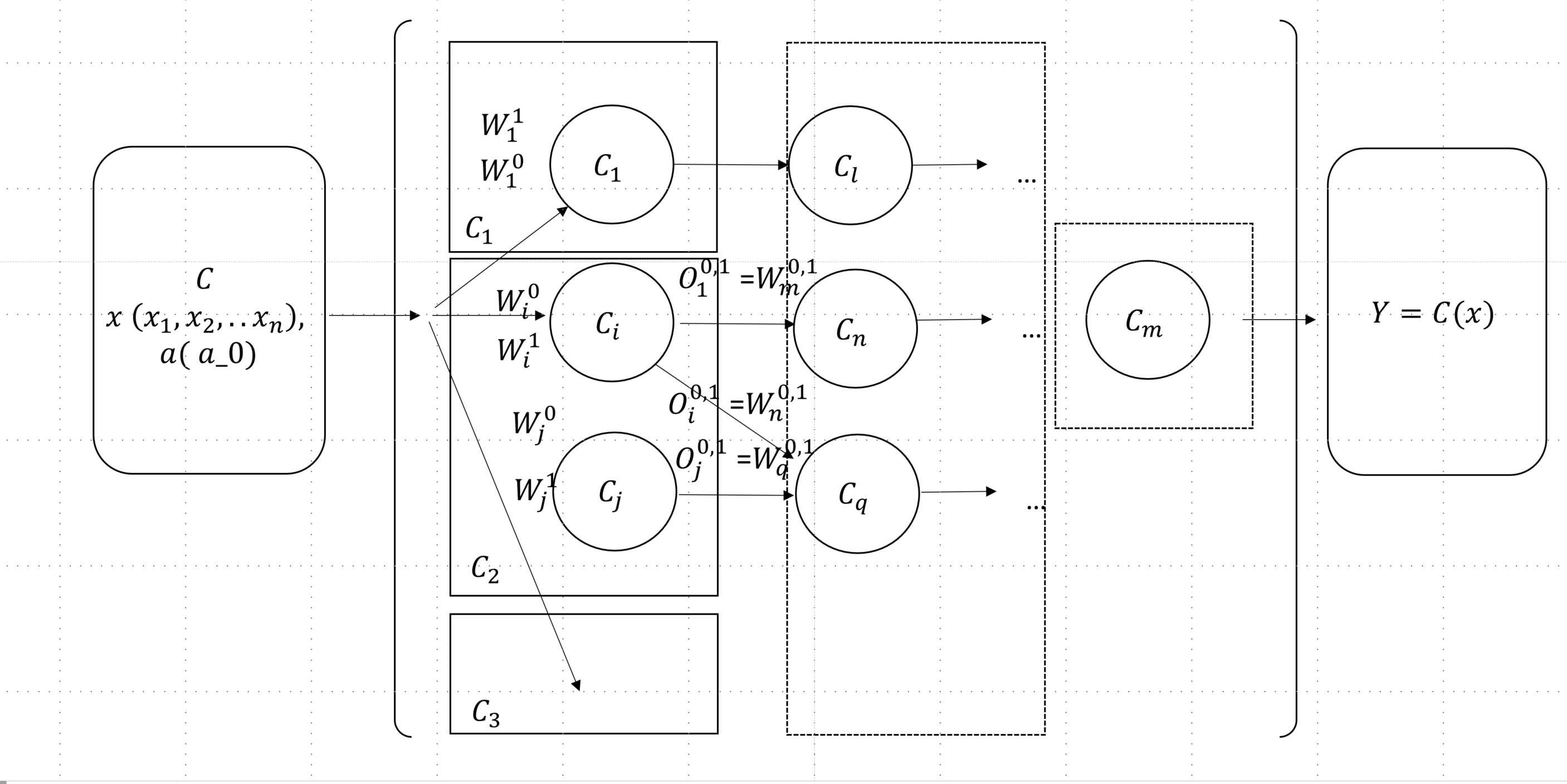}\hspace{2pc}%
\caption{Partitioned Garbled Circuit }
\end{figure}

\item[6] Partitioning Garbled Circuit: we can add the partitioned garbled circuit $(C_1, C_2,...C_m)$ and run the iterations of garbled circuit protocol per circuit.

\subitem[6.1] this sub-step runs in iterations.  Alice (prover) runs the Non-interactive multiple parties OT transfer scheme per partitioned circuit with multiple verifiers  offline to obtain the partitioned garbled circuit verification $Y_i = C_i(x_i,x_j)$,   except the last circuit $C_m$ introduced in section 5.3.

\item[7] for circuit $C_m$, it's required to employ the module III $OT-aggregator$ introduced in section 6 to obtain the combined Oblivious Transfer verification $Y= C_m(x_m)= \bigcup_{i=1}^{m-1} C_i(x_i,x_j)$.

\end{itemize}

\subsection{Offline Non-Interactive OT Transfer Protocol}
An offline non-interactive OT transfer protocol for the partitioned garbled circuit is illustrated in figure 6.
\begin{figure}[ht]
\centering
\includegraphics[width=0.9\columnwidth]{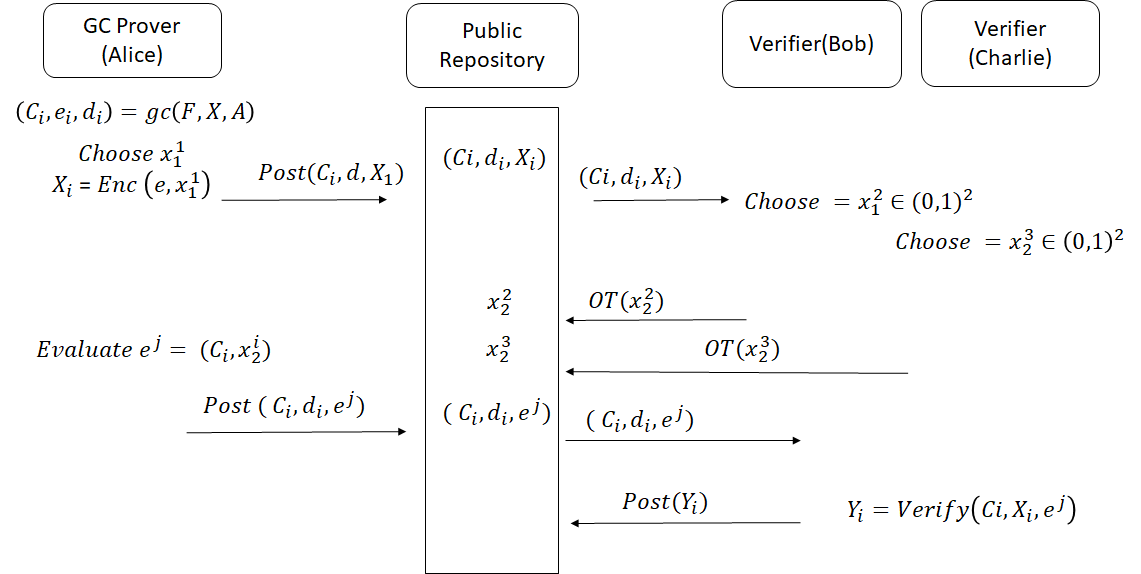}\hspace{2pc}%
\caption{Non-interactive Garbled Circuit  protocol }
\end{figure}

\begin{itemize}
\item[1] Alice represents the function $\hat{C}_i$ as a circuit and garbles circuit $C_i$. $\hat{C}_i$ has a total of 2 input wires corresponding to $(x_i,x_j)$.
\item[2] Alice sends over all of the ciphertexts that are generated for garbled circuit $C_i$ to a public repository, such as DLT in blockchain, or a public accessible web portal.
\item[3] Through the verification assignment system (not covered in this paper), the versifiers Bob, Charlie... commits to each of the garbled circuit verification, by exchanging the randomly generated number $x_i^2$ in the OT commitment message OT with Alice. We are employing the OT commitment scheme \cite{c20}  in our design. 
\item[4] Alice sends over the corresponding keys for its own inputs wire $x_i^1$.
\item[5] Alice and Bob, Charlie... in  OT oblivious transfer protocol introduced in section 7.A.
\item[6] At end of the protocol, Bob, Charlie,... learns the keys $d_i^j$ for each partitioned gates, they individually starts to evaluate the circuit using the keys obtained in the previous procedure.
\item[7] In the end,Bob, Charlie,... learns the output of $C(x_i,x_j)$. Alice also learns that Bob,Charlie,... has evaluated the results with outputs of $Y_1,Y_2,...$. if $Y_i \neq y_i$, Alice decides to abort, otherwise, Alice will proceed to Module III for OT aggregation.
\end{itemize}

\section{Module III: OT-aggregator Protocol}
The final stage of the zk-Fabric ends up with a OT-aggregator OT protocol  which combines all the partitioned garbled circuit verification in previous steps into an overall verification conclusion, as illustrated in figure 7. In most cases, this step involve the garbled circuit $C_m$ is built with the XOR gate in order to save the computational cost. The expression of the  circuit $C_m$  is 
    $Y= C_m(\bigcup_{i}^{m-1} Y_i\oplus X_1^{m})$.  The goal is for both prover (Alice) and aggregator (David) agree on the computed $Y=y$, where $y=f(x)$. 

\begin{figure}[ht]
\centering
\includegraphics[width=0.9\columnwidth]{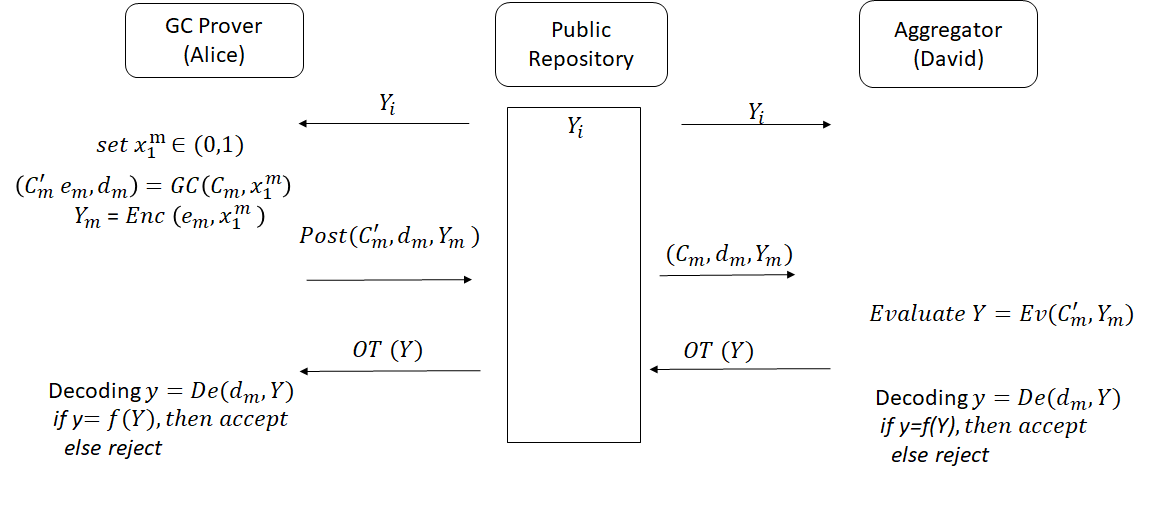}\hspace{2pc}
\caption{OT-aggregator protocol }
\end{figure}

\begin{itemize}

\item[1] At Module II, verifiers post the computed results $Y_i$ over the pubic repository where Alice and David can fetch from the portal. 
\item[2] Alice creates a random bits $x_1^m$, she lets the $x_1^m =(x_1^1 \oplus x_2^1)\oplus(x_1^2 \oplus x_2^2)\oplus ...)$.
\item[3] Alice garbles the circuit using the garbling algorithm. The garbling algorithm outputs $(C_m^{'}, e_m, d_m) \gets gc(C_m, x_1^m)$, where $gc()$ denotes a Yao's garbled circuit generation algorithm, and the $C_m=\bigcup_i^{m-1}(Y_i\oplus X_1^m)$. The output consists of a garbled circuit $C_m^{'}$, and encoding function $e_m$, and a decoding function $d_m$.
\item[4] Alice executes the deterministic encoding algorithm $Enc()$, which transforms $e_m$ and $x_1^m $ into the garbled input $Y_m=Enc(e_m, x_1^m)$. 
\item[5] Alice sents the turple  $(C_m^{'},d_m, Y_m)$ to the public repository, i.e the DLT in blockchain, or Web portal, where David is able to non-interactively fetch the information. 
\item[6] David creates a random bit $x_m^m \in (0,1)$.

\item[7] David also executes the deterministic evaluation algorithm $Ev()$, which outputs the $Y=Ev(C_m^{'}, x_m^m)$.

\item[8] David sents back the $Y$ to Alice through the public repository.

\item[9] Alice executes the deterministic decoding algorithm $De()$ to compute the final output $y=De(d_m,Y)$, where the $d_m$ denotes the decoding key. 

\subitem[9.1] - In parallel, David also executes the deterministic decoidng algorithm $De()$ to compute the final output $y=De(d_m,Y)$.

\item[10] At both Alice and David's sites, they will check if $y=f(m \oplus x_m^m)$ which $f()$ is the logical Boolean function before becoming garbled circuit function $C()$. If $yes$, the OT aggregator protocol will accept the verification results, otherwise, Alice should abort the verification.

\end{itemize}

\section{Conclusion and Future work}
The paper concludes by proposing a novel zero knowledge proof system which can handle more complex semantics. The construction of the zk-Fabric fits pretty well with distributed computing environments, such as Blockchain. In this paper, we have outlined the  modules and algorithms on section 3 to section 7 to support the overall functionalities of the zk-Fabric. 

Our future research should be focused on the prototyping the zk-Fabric and implement it in a testing environments. We need to measure the computational cost and resource cost comparing with other similar technologies. We should also further develop and  provide theoretical analysis of the security properties of the zk-Fabric schemes.


\bibliographystyle{ieeetr}
\bibliography{bibliography}

\end{document}